\documentclass[pre,twocolumn,showpacs]{revtex4}

\usepackage{graphicx}
\usepackage{bm}
\usepackage{amssymb}

\begin{document}
\title{Fractional oscillator}
\author{A. A. Stanislavsky}
\email{alexstan@ri.kharkov.ua} \affiliation{Institute of Radio
Astronomy, 4 Chervonopraporna~St., Kharkov 61002, Ukraine}

\date{\today}

\begin{abstract}
We consider the fractional oscillator being a generalization of
the conventional linear oscillator in the framework of fractional
calculus. It is interpreted as an ensemble average of ordinary
harmonic oscillators governed by stochastic time arrow. The
intrinsic absorption of the fractional oscillator results from the
full contribution of the harmonic oscillators' ensemble: these
oscillators differs a little from each other in frequency so that
each response is compensated by an antiphase response of another
harmonic oscillator. This allows to draw a parallel in the
dispersion analysis for the media described by the fractional
oscillator and the ensemble of ordinary harmonic oscillators with
damping. The features of this analysis are discussed.
\end{abstract}
\pacs{05.40.-a, 05.60.-k, 05.40.Fb} \maketitle

\section{Introduction}
The harmonic oscillator, given by a linear differential equation
of second order with constant coefficients, is a cornerstone of
the classical mechanics (see, for example, \cite{1,2}). Today this
elementary (and fundamental) conception has the widest origin of
physical, chemical, engineering applications and needs no
introduction. Its success mainly rests on its universality, and
its simplicity gives boundless intrinsic capabilities for sweeping
generalization. Suffice it to recall the passage from the language
of functions in phase space to operators in Hilbert space so that
the oscillatory model came strongly in the quantum theory
\cite{3,4}. Therefore no wonder, the fractional calculus has made
also an important contribution to this way.

At first the approach had a formal character by changing the
second derivative in the harmonic oscillator equation on the
derivative of an arbitrary order. After finding out the solutions
of such equations their relaxation-oscillation behavior was
established \cite{5,6}. The next step was a consideration of the
total energy and the phase plane representation for the fractional
oscillator \cite{7}. To save the dimension of energy, it is
necessary to generalize to the notation of momentum, though then
the parameter $m$ loses also the ordinary dimension of mass
\cite{8}. In this case the momentum is expressed in terms of the
Caputo-type fractional derivative \cite{6}. The fractional
oscillator is like a harmonic oscillator subject to a damping. The
source of the intrinsic damping is very intriguing. It is not
evident from fractional calculus, from the generalization of
derivative. The question requires an additional study exceeding
the bounds of fractional calculus itself.

Since it is a matter of the fractional integral/derivative with
respect to time, the answer to the aforementioned problem should
be sought by way of their concrete interpretation. Recently, the
probability interpretation of the temporal fractional
integral/derivative was suggested in \cite{9}. There exists a
direct connection between stable distributions in probability
theory and the fractional calculus. The occurrence of the temporal
fractional derivative (or integral) in kinetic equations indicates
the subordinated stochastic processes. Their directional process
is related to a stochastic process with a stable distribution. The
parameter characterizing the stable distribution coincides with
the index of the temporal fractional integral/derivative in the
corresponding kinetic equation. This means that such a equation
describes the evolution of a physical system whose time degree of
freedom becomes stochastic \cite{10}. The purpose of this paper is
to expand the interpretation on the fractional oscillator.

The paper is organized as follows. In Sec. \ref{par2} we analyze
an ensemble of harmonic oscillators with the stochastic time
clock. The new clock (random process) substitutes for the
deterministic time clock of the ordinary harmonic oscillator. The
nondecreasing random process arises from a self-similar
$\alpha$-stable random process of temporal steps. Using properties
of the stochastic time clock, we obtain the equation for the
fractional oscillator. In the spirit of this approach the
fractional oscillator can be considered as an ensemble average of
oscillators. Sec.~\ref{par3} is devoted to the comparison of
dispersion properties of the two media. One of them consists of
damped noninteracting harmonic oscillators, whereas another is the
fractional oscillator. It turns out that their dispersion
characteristics have a lot of common features. We discuss them in
detail. Our conclusions are briefly summarized in Sec.~\ref{par4}.
Appendix contains calculations for the response of the driven
fractional oscillator. They are useful for the dispersion analysis
in Sec.~\ref{par3}.

\section{Normal modes}\label{par2}
We start our consideration with the classical case of harmonic
oscillator. Based on the Hamilton function
$\mathcal{H}=(p^2+\omega^2q^2)/2$, where $p$ and $q$ are the
momentum and the coordinate, respectively, and $\omega$ the proper
frequency, the motion equations take the form
\begin{equation}
\partial q/\partial t=\partial\mathcal{H}/\partial p=p,\quad
\partial p/\partial t=-\,\partial\mathcal{H}/\partial q=-\omega^2.
\label{eq1}
\end{equation}
To multiply the first equation of (\ref{eq1}) on $\pm i\omega$ and
to add it with the second equation, we arrive at
\begin{equation}
\partial c/\partial t=-i\omega c,\qquad \partial c^*/\partial t=
i\omega c^*\,, \label{eq2}
\end{equation}
where the complex-conjugate values $c$ and $c^*$ satisfy the
relations
\begin{displaymath}
c=(\omega q+ip)/\sqrt{2\omega},\qquad c^*=(\omega q-i
p)/\sqrt{2\omega}.
\end{displaymath}
The solutions of Eqs.(\ref{eq2}) can be written as
\begin{eqnarray}
c(t)&=&c(0)e^{-i\omega t}=\frac{1}{\sqrt{2\omega}}[\omega
q(0)+ip(0)]e^{-i\omega t},\nonumber\\ c^*(t)&=&c^*(0)e^{i\omega
t}=\frac{1}{\sqrt{2\omega}}[\omega q(0)-ip(0)]e^{i\omega
t}.\label{eq4}
\end{eqnarray}
The values $c$ and $c^*$ are called else the normal modes of
oscillator \cite{1}. They have a very pictorial presentation in
the form of a vector rotating just as, the hand revolves around
the clock-face center with the frequency $\omega$.

A physical system of harmonic oscillators coupled to an
environment will interact with the environmental degrees of
freedom. This leads to a damping of oscillatory motion. If the
interaction manifests itself at random fashion, one of possible
ways to account for perturbations induced by the environment may
be as following. Let us randomize the time clock of the value
$c(\tau)$ so that any characteristic time is absent. Assume that
the time variable is a sum of random temporal intervals $T_i$ on
the non-negative semi-axis. If they are independent identically
distributed variables belonging to the strict domain of attraction
of a $\alpha$-stable distribution ($0<\alpha<1$), their sum has
asymptotically (the number of the intervals tends to infinity) the
stable distribution with the index $\alpha$. Following the
arguments of \cite{10,10a}, a new time clock is defined as a
continuous limit of the discrete counting process
$N_t=\max\{n\in{\bf N}: \sum_{i=1}^nT_i\leq t\}$, where ${\bf N}$
is the set of natural numbers. The time clock becomes the hitting
time process $S(t)$. Its basic properties is represented in
\cite{10a,11}. The probability density of the process $S(t)$ is
written in the form
\begin{equation}
p^{S}(t,\tau)=\frac{1}{2\pi i}\int_{Br} e^{ut-\tau
u^\alpha}\,u^{\alpha-1}\,du\,, \label{eq5}
\end{equation}
where $Br$ denotes the Bromwich path. This probability density has
a clear physical sense. It describes the probability to be at the
internal time $\tau$ on the real time $t$. In the case we
determine new normal modes
\begin{eqnarray}
c_\alpha(t)&=&\int_0^\infty
p^S(t,\tau)\,c(\tau)\,d\tau\,,\nonumber\\
c^*_\alpha(t)&=&\int_0^\infty p^S(t,\tau)\,\,c^*(\tau)\,d\tau\,.
\nonumber
\end{eqnarray}
The direct calculations give
\begin{eqnarray}
q_\alpha(t)&=&[c^*_\alpha(t)+c_\alpha(t)]/\sqrt{2\omega}=q(0)A(t)+
\frac{p(0)}{\omega}B(t)\,,\nonumber\\
p_\alpha(t)&=&i[c^*_\alpha(t)-c_\alpha(t)]\sqrt{\omega/2}=
p(0)A(t)+\omega q(0)B(t)\,,\nonumber
\end{eqnarray}
where
\begin{eqnarray}
A(t)&=&\int_0^\infty p^{S}(t,\tau)\,\cos\omega\tau\,d\tau=
E_{2\alpha,\,1}(-\omega^2t^{2\alpha})\,,\nonumber\\
B(t)&=&\int_0^\infty p^{S}(t,\tau)\,\sin\omega\tau\,d\tau=\omega
t^\alpha E_{2\alpha,\,\alpha+1}(-\omega^2t^{2\alpha})\,,\nonumber
\end{eqnarray}
and
\begin{displaymath}
E_{\mu,\,\nu}(z)=\frac{1}{2\pi
i}\int_Ce^u\,\frac{u^{\mu-\nu}\,du}{(u^\mu-z)}
\end{displaymath}
is the two-parameter Mittag-Leffler function \cite{12}. Here it is
easy to recognize the classical solutions for $\alpha=1/2$
(exponential function) and $\alpha=1$ (sine and cosine). The
functions $A(t)$ and $B(t)$ exhibit clearly the relaxation
features for $0<\alpha<1/2$, whereas for $1/2<\alpha<1$ the
functions represent a damping oscillatory motion. The latter case
just corresponds to the fractional oscillator. In particular the
value $A(t)$ satisfies the equation
\begin{displaymath}
A(t)=A(0)-\frac{\omega^2}{\Gamma(2\alpha)}\int^t_0(t-t')^{2\alpha-1}\,
A(t')\,dt'\,,
\end{displaymath}
with $A(0)=1$, where $\Gamma(x)$ denotes the gamma function. The
appropriate equation can be written also for $B(t)$. It should be
recalled here that the power kernel of fractional integral of the
order $\alpha$, $0<\alpha<1$ ``interpolates'' the memory function
between the Dirac $\delta$-function (the absence of memory) and
step function (complete ideal memory). This means that such memory
manifests itself within all the time interval $(0,t)$, but not at
each point of time (complete but not ideal memory). Under the
ideal complete memory the system ``remembers'' all its states, and
this excites the harmonic oscillations in such a system. The
absence of memory causes only the relaxation. The order of
fractional integral represents a quantitative measure of memory
effects \cite{12a}. In accordance with the theory of memory
effects the fractional oscillator contains simultaneously the
oscillatory motion and the relaxation.

From the series representation of $E_{\mu,\,\nu}(z)$ we derive the
leading asymptotic behavior of the value $A(t)$ and $B(t)$ for
$t\to 0$: $\lim_{t\to 0}A(t)=1, \lim_{t\to 0}B(t)=0$. According to
\cite{12}, the two-parameter Mittag-Leffler function approaches
zero as $z\to\infty$ in the sector of angles
$|\arg(-z)|<(1-\mu/2)\pi$, and increases indefinitely as
$z\to\infty$ outside of this sector. In our case we can use the
following expansion valid on the real negative axis
\begin{displaymath}
E_{\mu,\,\nu}(z)=-\sum_{n=1}^{N-1}\frac{z^{-n}}{\Gamma(\nu-n\mu)}
+ O(|z|^{-N}),\quad z\to -\infty\,.
\end{displaymath}
Thus, for $0<\alpha<1/2$ and $1/2<\alpha<1$ the value $A(t)$ and
$B(t)$ decreases algebraically in time.  As distinct from the case
of a damping harmonic oscillator, the model describes another
damping mechanism, without any external frictional force. The
damping of a fractional oscillator is due to internal causes
\cite{12b}. How to explain the attenuated oscillations? This
important feature of fractional oscillator has been already noted
from time to time in various publications \cite{5,7,8}. However,
the source of such intrinsic damping remained undecided.

We suggest the following interpretation. The fractional oscillator
should be considered as an ensemble average of harmonic
oscillators. When all harmonic oscillators are identical, and we
set their going in the same phase, their full contribution will be
equal to the product of the number of oscillators and the response
of one oscillator. This occasion appears if $\alpha=1$. However,
if the oscillators differ a little from each other in frequency,
even if they start in phase, after a while the oscillators are
allocated uniformly up to the clock-face. Each response will have
an antiphase response of another oscillator so that the total
response of all harmonic oscillators in such a system is
compensated. Although each oscillator is conservative (its total
energy saves), the system of such oscillators, resulting in the
fractional oscillator ($1/2<\alpha<1$), shows a dissipative
nature. In this connection it should be pointed out that the
similar situation may be observed also in the medium of harmonic
oscillators, having a given probability density on frequency (for
example, the Lorentz distribution \cite{13}). Both these cases are
closely connected with each other and have a common ground,
though, generally speaking, they describe different physical
systems. As has been shown in \cite{14,15}, Lagrangian and
Hamiltonian mechanics formulated with fractional derivatives in
time can be used for the description of nonconservative forces
such as friction. It should be also mentioned an interpretation of
fractional oscillator in \cite{15a}. In this case the Liouville
equation is formulated from a fractional analog of the
normalization condition for distribution function that can be
considered in a fractional phase space. The latter has a
fractional dimension as well as the fractional measure. The volume
element of the fractional phase space is realized by fractional
exterior derivatives. The usual  nondissipative systems become
dissipative in the fractional phase space. However, the approach
is different on ours. It operates with fractional powers of
coordinates and momenta. Such fractional systems are nonlinear.

\section{Dispersion}\label{par3}
Now we examine the behavior of the fractional oscillator under the
influence of an external force. From above this case corresponds
to oscillations in the ensemble of nonidentical harmonic
oscillators noninteracting with each other. In the framework of
this model the fractional oscillator with the initial conditions
$x(0)=0$ and $\dot x(0)=0$ is described by the following equation
\begin{eqnarray}
x(t)&=&-\frac{\omega^\alpha_0}{\Gamma(\alpha)}\int^t_0(t-t')^{\alpha-1}\,
x(t')\,dt'+\nonumber\\
&+&\frac{1}{\Gamma(\alpha)}\int^t_0(t-t')^{\alpha-1}\,
F(t')\,dt'\,,\label{eq6}
\end{eqnarray}
where it should be kept $1<\alpha<2$, and $F$ is the external
force. The dynamic response of the driven fractional oscillator
was investigated in \cite{8}:
\begin{equation}
x(t)=\int^t_0F(t')\,(t-t')^{\alpha-1}\,E_{\alpha,\,\alpha}
(-\omega^\alpha_0(t-t')^\alpha)\,dt'\,. \label{eq7}
\end{equation}
This allows us to define the response for any desired forcing
function $F(t)$. The ``free'' and ``forced'' oscillations of a
fractional oscillator depend on the index $\alpha$. However, in
the first case the damping is characterized only by the ``natural
frequency'' $\omega_0$, whereas the damping in the case of
``forced'' oscillations depends also on the driving frequency
$\omega$. Each of these cases has a characteristic algebraic tail
itself, associated with damping \cite{12b}.

Let $F(t)$ be periodic, $F(t)=F_0\,e^{j\omega t}$. Then the
solution of Eq. (\ref{eq6}) is determined by taking the inverse
Laplace transform
\begin{equation}
x(t)=\frac{1}{2\pi
j}\int_{Br}e^{st}\frac{F_0\,(s+j\omega)\,ds}{(s^2+\omega^2)\,
(s^\alpha+\omega_0^\alpha)}\,.\label{eq8}
\end{equation}
The Bromwich integral (\ref{eq8}) can be evaluated in terms of the
theory of complex variables. Some particular examples of forcing
functions were considered in \cite{8}. However, the set turns out
to be enough scanty for our aim. The necessary computations with
$F(t)=A\sin(\omega t+\phi)$ are fulfilled in Appendix. The phase
$\phi$ is constant.

If one waits for a long enough time, the normal mode of this
system is damped. Therefore, consider only the forced oscillation.
After the substitution of $\bar x(t)=x_0e^{j\omega t}$ for $x(t)$
in Eq.(\ref{eq6}) we obtain
\begin{eqnarray}
x_0\,e^{j\omega t}
&=&-\frac{\omega^\alpha_0}{\Gamma(\alpha)}\int^t_0(t-t')^{\alpha-1}\,
x_0\,e^{j\omega t'}\,dt'+\nonumber\\
&+&\frac{1}{\Gamma(\alpha)}\int^t_0(t-t')^{\alpha-1}\,
F_0\,e^{j\omega t'}\,dt'\,.\label{eq9}
\end{eqnarray}
It is convenient to change the variable $\omega(t-t')=\zeta$ in
the integrand. Next we can divide out $\exp(j\omega t)$ from each
side of (\ref{eq9}) and direct $t$ to infinity. The procedure
permits to extract the contribution of steady-state oscillations.
Using the table integral \cite{16}
\begin{displaymath}
\int_0^\infty
z^{\alpha-1}\,e^{-jz}\,dz=\Gamma(\alpha)\,e^{-j\pi\alpha/2}\,,
\end{displaymath}
Eq. (\ref{eq9}) gives
\begin{equation}
x_0=\frac{F_0}{[\omega_0^\alpha+\omega^\alpha\exp(j\pi\alpha/2)]}
\,.\label{eq10}
\end{equation}
This result is completely confirmed by a more comprehensive
analysis given in Appendix.

\begin{figure}
\includegraphics[width=8.6 cm]{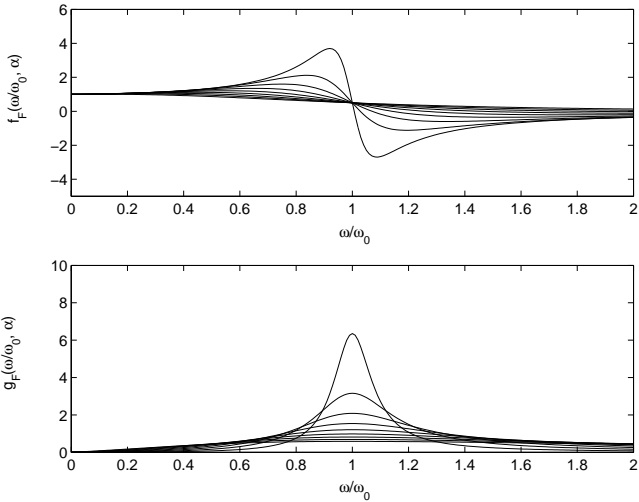}
\caption{\label{fig1}Dispersion dependence of the fractional
oscillator in the form of the functions
$f_F(\omega/\omega_0,\alpha)$ and $g_F(\omega/\omega_0,\alpha)$
with different values $\alpha$, from 0.1 to 0.9 with the step
0.1.}
\end{figure}

As is well know, the ensemble behavior of identical noninteracting
harmonic oscillators is a basic topic for considering in the
classical theory of dispersion. The nonidentity of oscillators is
necessary to take into account, for example, for the dispersion
analysis of propagating electromagnetic waves into a heated gas,
where the spread in molecule velocity values leads to a Doppler
shift of the oscillators' normal frequency with respect to the
forced field frequency. Right now let a medium of oscillators be
such that results in the fractional oscillator. It interests us
the polarizability of such a medium.  In this case the
permittivity is written as $\epsilon=1+4\pi e^2x_0/(F_0\,m)$,
where $e$ is the electron charge. It should be pointed out that in
contrast to a simple harmonic oscillator the parameter $m$ does
not have the ordinary dimension of mass. However, the generalized
momentum $p$ of the fractional oscillator is defined via the
Caputo-type fractional derivative of order $\alpha/2$ \cite{6} so
that the expression $p^2/(2m)$ has the dimension of energy (see
details in \cite{8}). The real and imaginary parts of the
permittivity take the form
\begin{eqnarray}
\mathrm{Re}\,\epsilon(\omega)=1+\frac{4\pi e^2
[\omega_0^\alpha+\omega^\alpha\,
\cos(\pi\alpha/2)]}{m\,[\omega_0^{2\alpha}+\omega^{2\alpha}+2\omega_0^\alpha
\omega^\alpha\cos(\pi\alpha/2)]}\,,\label{eq11}\\ \mathrm{Im}\,
\epsilon(\omega)=-\frac{4\pi e^2\omega^\alpha\,
\sin(\pi\alpha/2)}{m\,[\omega_0^{2\alpha}+\omega^{2\alpha}+2\omega_0^\alpha
\omega^\alpha\cos(\pi\alpha/2)]}\,.\label{eq12}
\end{eqnarray}
For $\alpha=2$ we arrive at the Sellmeier's formula \cite{17}:
\begin{displaymath}
\epsilon(\omega)=1+4\pi e^2N/[m(\omega_0^2-\omega^2)]\,,
\end{displaymath}
where we include $N$ to account for the number of harmonic
oscillators in the medium. In this case the parameter $m$ is
really the electron mass. The index $\alpha=2$ corresponds to the
classical harmonic oscillator without any damping, and all the
oscillators in the ensemble go in the same phase. Therefore, the
Sellmeier's formula contains only the real part of permittivity.

\begin{figure}
\includegraphics[width=8.6 cm]{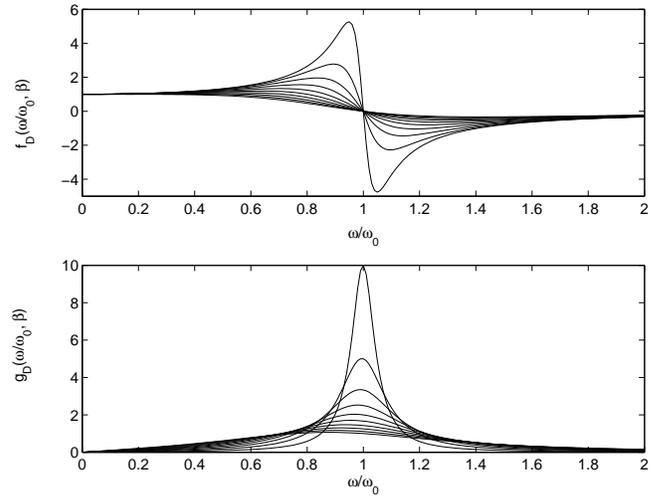}
\caption{\label{fig2}Dispersion dependence in the classical case
(ensemble of ordinary harmonic oscillators with damping) with
different values $\beta$, from 0.1 to 1.0 with the step 0.1.}
\end{figure}

We can conduct a clear comparison between the dispersion
characteristics of the fractional oscillator and one of an
ensemble of classical harmonic oscillators with damping. Normalize
the frequency $\omega$ in their permittivity by $\omega_0$. In
fact the constants (like $e$, $m$ and so on) in
$\mathrm{Re}\,\epsilon(\omega)$ and
$\mathrm{Im}\,\epsilon(\omega)$ define only a scale. Thus, one can
pick out the functional dependence of these permittivities on
$\omega/\omega_0=z$. Denote $2\gamma/\omega_0$ by $\beta$, where
$\gamma$ defines the damping in each classical harmonic
oscillator. Then we have the following dependences for the
fractional oscillator
\begin{eqnarray}
\mathrm{Re}\,\epsilon_F(\omega)\quad\to\quad
f_F(z,\alpha)&=&\frac{1+z^\alpha\,
\cos(\pi\alpha/2)}{z^{2\alpha}+2z^\alpha\cos(\pi\alpha/2)+1}\,,
\nonumber\\ \mathrm{Im}\,\epsilon_F(\omega)\quad\to\quad
g_F(z,\alpha)&=&\frac{z^\alpha\sin(\pi\alpha/2)}{z^{2\alpha}+
2z^\alpha\cos(\pi\alpha/2)+1}\,,\nonumber
\end{eqnarray}
and ones for the classical harmonic oscillators with damping
\begin{eqnarray}
\mathrm{Re}\,\epsilon_D(\omega)\quad\to\quad
f_D(z,\beta)&=&\frac{1-z^2} {(1-z^2)^2+z^2\beta^2}\,,\nonumber\\
\mathrm{Im}\,\epsilon_D(\omega)\quad\to \quad g_D(z,\beta)&=&
\frac{z\beta}{(1-z^2)^2+z^2\beta^2}\,.\nonumber
\end{eqnarray}
If the parameter $\beta$ determines the damping value in the
harmonic oscillator, the index $\alpha$ just characterizes the
same for the fractional oscillator. The extremum values of
$f_D(z,\beta)$ and $g_D(z,\beta)$ decrease with increasing the
parameter $\beta$ whereas for $f_F(z,\alpha)$ and $g_F(z,\alpha)$
vice versa: the extremum values increase with increasing the index
$\alpha$, though it should be noted that this index itself belongs
only to the interval $1<\alpha\leq 2$. The functions
$f_{\cdots}(z)$ and $g_{\cdots}(z)$ are shown on Fig.~\ref{fig2}
and Fig.~\ref{fig3}.

From the relations (\ref{eq11}) and (\ref{eq12}) it follows that
there is a frequency range, where the absorption is small, and the
refraction coefficient increases with frequency (normal
dispersion). Moreover, in the frequency range, where the
absorption is big, the anomalous dispersion happens to be the case
for the refraction coefficient decreasing with frequency. In this
connection it should be pointed out that the presence of the
normal and anomalous  dispersion is typical for such an ensemble
of ordinary harmonic oscillators and is well known. However a new
fact established here is that the normal and anomalous dispersion
is also typical for the medium described as a fractional
oscillator.

\section{Summary}\label{par4}
We have shown that the fractional oscillator can be considered as
a model of the harmonic oscillators' medium. Its stochastic
properties accumulate in the index of the fractional
integral/derivative with respect to time. The frequency
distinction of the oscillators (constituents of the fractional
oscillator) from each other is at the bottom of the intrinsic
damping for such a system. As a consequence, the dispersion
properties of the medium, like the fractional oscillator, is
enough similar to the case, when a medium is modeled by an
ensemble of harmonic oscillators with damping.

\section*{Appendix}
\begin{figure}
\includegraphics[width=8.6 cm]{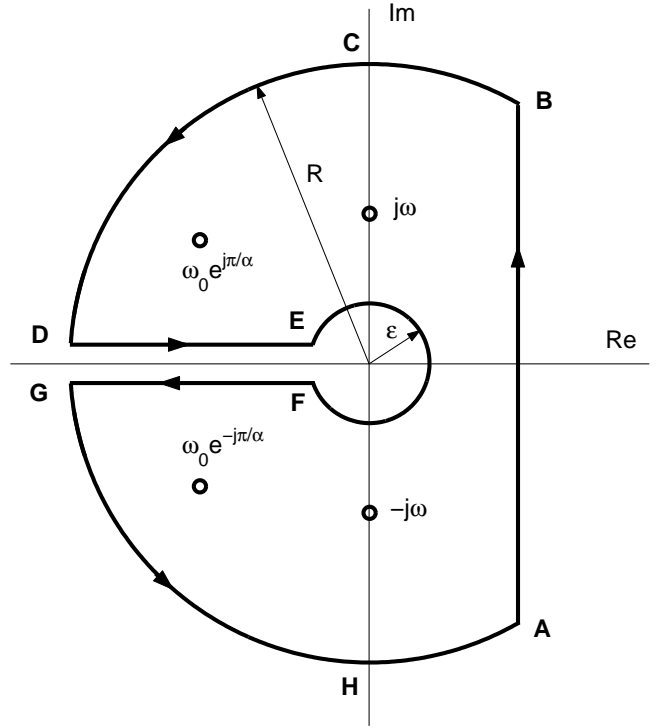}
\caption{\label{fig3}Contour, inside which the function $\tilde
x(s)$ remains single-valued and analytical all over, with the
exception of poles $\pm j\omega$ and $\omega_0\exp(\pm
j\pi/\alpha)$.}
\end{figure}

We here derive properties of the response function (\ref{eq7}) for
the forcing function $A\sin(\omega t+\phi)$ directly from its
representation as a Laplace inverse integral
\begin{eqnarray}
x(t)&=&\frac{1}{2\pi j}\int_{Br}e^{st}\,\tilde
x(s)\,ds=\nonumber\\ &=&\frac{1}{2\pi
j}\int_{Br}e^{st}\frac{A\,(s\cdot\sin\phi+\omega\cdot\cos\phi)\,ds}
{(s^2+\omega^2)\, (s^\alpha+\omega_0^\alpha)}\,,\quad (\mathrm
{A1})\nonumber
\end{eqnarray}
where the phase $\phi$ is constant, ${\it Br}$ denotes the
Bromwich path, and $1<\alpha\leq~2$. By bending the Bromwich path
into the equivalent Hankel path (Fig.~\ref{fig3}), the response
function $x(t)$ can be decomposed into two contributions.

The first contribution arises from the two borders of the cut
negative real axis (lines {\bf DE} and {\bf FG}):
\begin{displaymath}
x_1(t)=-\frac{1}{2\pi j}\int^0_{-\infty}e^{st}\tilde
x(s)\,ds-\frac{1}{2\pi j}\int^{-\infty}_0e^{st}\tilde x(s)\,ds.
\end{displaymath}
To enter $s=re^{j\pi}$ into the integral taken along the upper
border and $s=re^{-j\pi}$ into the integral along the lower
border, we get
\begin{displaymath}
x_1(t)=\int_0^\infty e^{-rt}M_\alpha(r)\,dr
\end{displaymath}
with
\begin{displaymath}
M_\alpha(r)=\frac{A\,r^\alpha\,(\omega\,
\cos\phi-r\,\sin\phi)\,\sin\pi\alpha}{\pi\,(r^2+\omega^2)\,(r^{2\alpha}+2r^\alpha
\omega_0^\alpha\cos\pi\alpha+\omega_0^{2\alpha})}.
\end{displaymath}

The second contribution is determined by the Cauchy theorem on
residues. The integrand of (A1) has the following poles
\begin{displaymath}
s=\pm j\omega\qquad\mathrm{and}\qquad s=\omega_0e^{\pm
j\pi/\alpha}.
\end{displaymath}
Calculating the residues of the poles $s=\pm j\omega$, we obtain
\begin{displaymath}
x_2'(t)=A\left[\frac{\omega_0^\alpha\,\sin(\omega
t+\phi)+\omega^\alpha\sin(\omega t-\pi\alpha/2+\phi)}
{\omega_0^{2\alpha}+\omega^{2\alpha}+2\omega_0^\alpha
\omega^\alpha\cos(\pi\alpha/2)}\right].
\end{displaymath}
It remains to define the residues for the two other poles:
\begin{eqnarray}
&&\left[\frac{e^{st}(s\cdot\sin\phi+\omega\cdot\cos\phi)}
{(s^2+\omega^2)\frac{d}{ds}(s^\alpha+\omega_0^\alpha)}
\right]_{s=\omega_0e^{\pm j\pi/\alpha}}=\nonumber\\ &&\nonumber\\
&=&\frac{e^{\omega_0t(\cos\pi/\alpha\pm
j\sin\pi/\alpha)}[\omega_0\,e^{\pm
j\pi/\alpha}\,\sin\phi+\omega\cos\phi]}{\alpha\,\omega^{\alpha-1}\,e^{\pm
j\pi(\alpha-1)/\alpha}\,[\omega_0^2 e^{\pm
2j\pi/\alpha}+\omega^2]}.\nonumber\\ &&\nonumber
\end{eqnarray}
They lead to
\begin{displaymath}
x_2''(t)=\frac{2
A\,\exp(\omega_0t\cos(\pi/\alpha))\,[C\cos\phi-D\sin\phi]}
{\alpha\,\omega^{\alpha-1}\,(\omega_0^4+\omega^4+2\omega_0^2\omega^2\cos
(2\pi/\alpha))},
\end{displaymath}
where
\begin{eqnarray}
C&=&\omega[\omega_0^2\cos(\omega_0t\sin(\pi/\alpha)-
\pi(1+\alpha)/\alpha)+\nonumber\\ &+&\omega^2
\cos(\omega_0t\sin(\pi/\alpha)+\pi(1-\alpha)/\alpha)],\nonumber\\
D&=&\omega_0[\omega_0^2\cos(\omega_0t\sin(\pi/\alpha))+\nonumber\\
&+&\omega^2\cos(\omega_0t \sin(\pi/\alpha)+2\pi/\alpha)].\nonumber
\end{eqnarray}

As a result, the response function $x(t)$ takes the form:
\begin{displaymath}
x(t)=x_1(t)+x_2'(t)+x_2''(t)\,.
\end{displaymath}
Since $\cos(\pi/\alpha)<0$, the term $x_2''(t)$ describes the
relaxation of the normal mode in this system. For $1<\alpha<2$ and
all $r$ the denominator of the value $M_\alpha(r,\alpha)$ is
always positive:
$(r^{2\alpha}+2r^\alpha\omega_0^\alpha\cos\pi\alpha+
\omega_0^{2\alpha})>(r^\alpha-\omega^\alpha_0)^2\geq 0$, and the
term $\sin\pi\alpha$ is always negative.  Depending on $\phi$,
each of terms $\omega\,\cos\phi$ and $r\,\sin\phi$ may be both
positive and negative. However the value $x_1(t)$ becomes
vanishingly small with $t\to\infty$. The steady-state oscillation
in this system is defined only by the term $x_2'(t)$. The latter
can be expressed as $x_2'(t)=A_1\sin(\omega t+\phi-\delta)$, where
\begin{eqnarray}
A_1&=&\frac{A}{[\omega^{2\alpha}+\omega_0^{2\alpha}+2\omega^\alpha
\omega_0^\alpha\cos(\pi\alpha/2)]^{1/2}},\nonumber\\
\delta&=&\arctan\left[\frac{\omega^\alpha\sin(\pi\alpha/2)}
{\omega^\alpha\cos(\pi\alpha/2)+\omega_0^\alpha}\right].\nonumber
\end{eqnarray}
To put $\phi=0$ in (A1), we arrive at the results of section 4.3
from \cite{8}. It should be also noted that the oscillatory
contribution $x_2''(t)|_{\phi=0}$ has some resemblance with the
``free'' oscillations of a damped harmonic oscillator and the
forced oscillations of a driven damped harmonic oscillator
\cite{12b}.

\bibliographystyle{apsrev}
\bibliography{oscillator}

\end{document}